\documentclass[usenatbib]{mn2e}

\usepackage[dvips]{graphicx}
\usepackage{amssymb}
\usepackage{txfonts}

\newcommand{\msun}{{\rm M}_{\sun}}

\newcommand{\g}{$\gamma$}

\title[Cyg X-3: a low-mass black hole or a neutron star]{Cyg X-3: a low-mass black hole or a neutron star}

\author[A. A. Zdziarski et al.]
{Andrzej A. Zdziarski,$^{1}$ Joanna Miko{\l}ajewska$^{1}$ and Krzysztof Belczy{\'n}ski$^{2,3}$\\
 $^{1}$Centrum Astronomiczne im.\ M. Kopernika, Bartycka 18,
 PL-00-716 Warszawa, Poland\\ 
 $^{2}$Astronomical Observatory, University of Warsaw, Al.
 Ujazdowskie 4, 00-478 Warsaw, Poland\\
 $^{3}$Center for Gravitational Wave Astronomy, University of Texas at
 Brownsville, Brownsville, TX 78520, USA
  }

\date{Accepted 2012 November 13. Received 2012 October 24; in original form 2012 August 27}

\pagerange{\pageref{firstpage}--\pageref{lastpage}}
\pubyear{2012}

\begin{document}

\maketitle

\label{firstpage}

\begin{abstract}
Cyg X-3 is a highly interesting accreting X-ray binary, emitting from the radio to high-energy gamma-rays. It consists of a compact object wind-fed by a Wolf-Rayet (WR) star, but the masses of the components and the mass-loss rate have been a subject of controversies. Here, we determine its masses, inclination, and the mass-loss rate using our derived relationship between the mass-loss rate and the mass for WR stars of the WN type, published infrared and X-ray data, and a relation between the mass-loss rate and the binary period derivative (observed to be $>0$ in Cyg X-3). Our obtained mass-loss rate is almost identical to that from two independent estimates and consistent with other ones, which strongly supports the validity of this solution. The found WR and compact object masses are $10.3_{-2.8}^{+3.9} \msun$, $2.4_{-1.1}^{+2.1} \msun$, respectively. Thus, our solution still allows for the presence of either a neutron star or a black hole, but the latter only with a low mass. However, the radio, infrared and X-ray properties of the system suggest that the compact object is a black hole. Such a low-mass black-hole could be formed via accretion-induced collapse or directly from a supernova.
\end{abstract}
\begin{keywords}
binaries: close -- stars: individual: Cyg~X-3 -- stars: winds, outflows -- stars: Wolf-Rayet -- X-rays: binaries.
\end{keywords}

\section{Introduction}
\label{intro}

Cyg X-3 is an X-ray binary possessing a number of unique and highly interesting characteristics. It is the brightest radio source among X-ray binaries \citep{mccollough99}, showing extremely strong radio outbursts and resolved jets (e.g., \citealt*{marti01,m01}). It is, so far, the only X-ray binary that is certainly powered by accretion for which emission of high-energy \g-rays has been unambiguously confirmed \citep{agile,fermi09}. It is also the only known X-ray binary in the Galaxy with a Wolf-Rayet (WR) donor \citep*{v92,v96,v93,fender99}. Furthermore, its orbital period of $P\simeq 0.2$ d is unusually short for a high-mass binary, which indicates a past spiral-in episode during a common-envelope evolutionary stage. Given the value of $P$ and estimates of the masses and mass-loss rate, the compact object orbits within the WR photosphere. 

In spite of the discovery of Cyg X-3 already in 1966 \citep{giacconi67}, the nature of its compact object has remained unknown; it may be either a neutron star (NS) or a black hole (BH, e.g., \citealt*{hanson00,vilhu09}, hereafter V09). This issue is of great interest since Cyg X-3 is a likely progenitor of a double compact system, after its WR star explodes as a supernova. If it contains an NS, it may be a progenitor of double NS systems like PSR B1913+16 \citep{ht75}. That pulsar has a relatively short spin period, 59 ms, and a weak magnetic field, $\simeq 2\times 10^{10}$ G. It might have been recycled by accretion in a system like Cyg X-3, which both weakened the initially strong magnetic field and spun it up before the donor exploded and formed a second NS (see, e.g., \citealt{vdh95} and references therein). On the other hand, if it contains a BH, it may lead to formation of a BH-NS or double BH system, which merger can then be detectable in gravitational waves \citep{paper2}. Evolutionary calculations and population synthesis for Cyg X-3 were performed by \citet{lommen05}. They found that either a BH or an NS could be present, although they favoured the BH. 

Here, we self-consistently determine the binary parameters and the mass-loss rate, $\dot M$, of the donor. We use published results for the radial velocities of the two components, and a relationship between the rate of the period increase, $\dot P$, $\dot M$, and the total mass of the system. We close the equations by finding a tight correlation between $\dot M$ and the WR mass, $M_{\rm WR}$ (for the WN stellar type, identified to be present in Cyg X-3).

\section{The binary parameters and mass-loss rate}
\label{masses}

The best existing determination of the mass function of the compact object appears to be that of \citet{hanson00}, who measured the radial velocity of the WR star as $K_{\rm WR}\simeq 109\pm 13$ km/s. Then, V09 studied orbital modulation of X-ray lines in the system using {\it Chandra}, in particular the lines originating close to the compact object. In their determination of the radial velocity of the compact object, $K_{\rm C}$, they calculated the average based on 4 and 10 bins. They also considered the latter case with the two bins having the maximum blueshift removed. That method, which we consider questionable, gave the radial velocity substantially less than the other ones, which reduced the resulting velocity average. Here, we use their values for the 4 and 10 bins with, in addition, the {\it Chandra\/} observation of Obsid 6601 from V09. The unweighted average of the three measurements gives then the compact-object velocity of $K_{\rm C}\simeq 469\pm 88$ km/s. We have also repeated their fits for their data from fig.\ 7, obtaining values very similar to their, and consistent with the above average. Thus, we use this value of $K_{\rm C}$ hereafter (although we find below that our conclusions change only slightly if the original value of V09 is used).

With the binary period of $P=0.19969$ d \citep{singh02}, the mass functions are $f_{\rm WR}= 2.13^{+1.44}_{-0.99} \msun$ and $f_{\rm C}= 0.027^{+0.011}_{-0.008}\msun$. (The original value of $K_{\rm C}\simeq 418\pm 123$ km/s of V09 implies $f_{\rm WR}= 1.51^{+1.59}_{-0.92} \msun$.) This (for our value of $f_{\rm WR}$) implies the mass ratio and the inclination of 
\begin{eqnarray}
\lefteqn{
q\equiv {M_{\rm C}\over M_{\rm WR}}={K_{\rm WR}\over K_{\rm C}} \simeq 0.23^{+0.09}_{-0.06},\label{ratio}}\\
\lefteqn{
\sin i =\left[(K_{\rm WR}+ K_{\rm C})^2 K_{\rm C} P\over 2\pi G M_{\rm WR}\right]^{1/3}
\simeq (0.69\pm 0.12) \left(M_{\rm WR}\over 10\msun\right)^{-1/3},
 \label{incl}}
\end{eqnarray}
respectively (note some misprints in the corresponding equations of V09). Here, we have estimated the errors as corresponding to the extrema of $K$. At $M_{\rm WR}=10\msun$, $i\simeq {43\degr}^{+11}_{-9}$. If $i\geq 30\degr$, which is likely given the strong IR and X-ray orbital modulation, the above constraints imply $M_{\rm WR}\la 42\msun$ and the compact-object mass of $M_{\rm C}\la 9\msun$. 

V09 self-consistently explain the radial velocity and flux modulations with the orbital phase for both the X-ray and near-IR features (lines and continuum) in the frame of a model consisting of a WR star and a compact companion. In particular, if the minimum flux (X-ray, near-IR) occurs when the WR transits in front of the compact object, the radial velocity curves tracing the WR star (He{\sc i} absorption; \citealt{hanson00}) and the compact object (Fe{\sc xxvi}; V09) have the minima/maxima at the respective quadratures, i.e., are shifted by $0.25 P$ and $0.75 P$, respectively, whereas the radial velocities of the IR He{\sc ii} emission formed in the X-ray shadow cone behind the WR star, where the wind is photo-ionized by its EUV radiation, show a direct correlation with the light curve\footnote{\citet{lommen05} also considered the case of an NS accreting at a high rate through a Roche-lobe overflow from a low-mass, $\la 1.5\msun$, He star. Such a star has a negligible wind and the observed WR features originate in an outflow centred on the NS, implying that both IR and X-ray spectral features originate close to the NS. It then appears impossible to explain the orbital modulation of the continuum, the emission lines fluxes and the radial velocities. Specifically, both the lines originating in the WR-like outflow and around the NS should have similar dependencies on the orbital phase, contrary to the observations (\citealt{hanson00}; V09). We note that this model was proposed before the results of V09 were obtained.}.

We have then two additional constraints. The binary is detached, see equations (\ref{rwr}--\ref{roche}) below, and thus there is no Roche lobe overflow and accretion is from the wind. Most of the wind leaves the system carrying away its angular momentum. This leads to slowing down the binary rotation at a rate (e.g., \citealt{do74}),
\begin{equation}
{\dot P\over P}\simeq {2\dot M\over M},
\label{pdot}
\end{equation}
where $M=M_{\rm WR}+M_{\rm C}$. \citet*{ogley01} critically considered assumptions made in obtaining this relationship. Based on their discussion, there appear to be no major problems with this method when applied to Cyg X-3. The most recent published estimate for $\dot P/P$ is $(1.05\pm 0.04) \times 10^{-6}$ yr$^{-1}$ \citep{singh02}\footnote{\citet{singh02} and Kitamoto also fitted cubic ephemerides to their data, with $\ddot P\neq 0$, obtaining rather short time scales of $\dot P/\ddot P\simeq -60\pm 70$ yr and $-130\pm 80$ yr, respectively. This would predict that $\dot P$ changes its sign in $\sim 10^2$ yr, which is $\ll$ any evolutionary time scale of the wind, making this model inconsistent with the $\dot P$ being due to the loss of the angular momentum via wind. If such an ephemeris were shown to be correct, another explanation, e.g., a third body, would be required; however, a linear + sinusoidal ephemeris gives a very bad fit to the data. Moreover, the cubic ephemeris for the most recent data set of Kitamoto yields no improvement of the fit over the parabolic one.}. It has been confirmed and improved by S. Kitamoto (private comm.) by including {\it Suzaku\/} observations, which yields $\dot P/P\simeq (1.03\pm 0.02) \times 10^{-6}$ yr$^{-1}$. Then, $M\simeq 2\dot M P/\dot P\simeq 19\msun (\dot M/10^{-5}\msun/{\rm yr})$. 

\begin{figure}
\centerline{\includegraphics[width=\columnwidth]{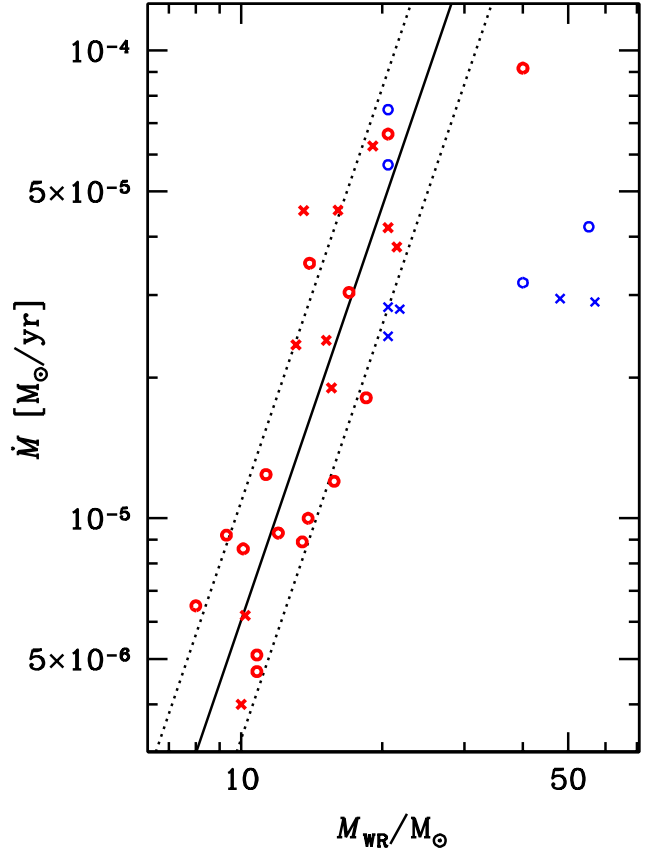}}
\caption{$\dot M$ corrected for wind clumping vs.\ $M_{\rm WR}$ for WN stars in the sample of NL00. The solid line shows the fit to all points with $M_{\rm WR} <22\msun$, and the dotted lines delineate the rms of these points. The crosses and circles show the single stars and those in binaries, respectively. We see no systematic difference between the two subsamples. The thick (red) and thin (blue) symbols show stars with $Y>0.8$ and $<0.8$, respectively.
}
\label{mdot_vs_m}
\end{figure}

The second constraint is of the mass-loss rate vs.\ mass relationship in WR stars. \citet{v96} classified Cyg X-3 as the WN 4--8 stellar type, which is intermediate between the WNE type, almost H free, and WNL, with substantial fraction of H, on the basis of the I and K-band emission lines. We use the sample of table 5 of Nugis \& Lamers (2000, hereafter NL00), containing 34 WN stars (31 of them with the type WN 4--8), and giving $\dot M$ corrected for wind clumping (unlike a number of previous studies), which we show in Fig.\ \ref{mdot_vs_m}. The metallicity of all these stars is close to solar, $Z= 0.0172$--0.0176. We show separately single stars (crosses, 15 stars) and those in binaries (circles, 19 stars), but see no systematic differences in $\dot M(M_{\rm WR})$ between the two classes. Also, we distinguish between the stars with low and high H content, showing in thick (red) and thin (blue) symbols stars with the He fraction of $Y>0.8$ and $Y<0.8$, respectively. The hydrostatic core radii for the former are \citep{sm92}
\begin{equation}
R_{\rm WR}\simeq 6\times 10^{10}\left(M_{\rm WR}\over 10\msun\right)^{0.584}{\rm cm},
\label{rwr}
\end{equation}
as adopted by NL00. The volume-averaged Roche-lobe radii are \citep{p71}
\begin{equation}
R_{\rm RL}\simeq \left[2 G M_{\rm WR} P^2\over (9\upi)^2\right]^{1/3}\simeq 1.0\times 10^{11}\left(M_{\rm WR}\over 10\msun\right)^{1/3}{\rm cm},
\label{roche}
\end{equation}
valid for $M_{\rm WR}/M\la 0.8$ (approximately consistent with our values of $q$). Thus, $R_{\rm WR}<R_{\rm RL}$ for WR stars with $Y>0.8$ at any mass of interest. On the other hand, stars with $Y<0.8$ have substantially larger radii and their presence in Cyg X-3 can be thus excluded. However, we see in Fig.\ \ref{mdot_vs_m} that although they have on average larger masses, $M_{\rm WR}\ga 20\msun$ (which is due to their earlier evolutionary status), than those with $Y>0.8$, they still follow the same $\dot M(M_{\rm WR})$ dependence. On the other hand, we see a major difference between stars with low masses, $M_{\rm WR}<22\msun$ (29 stars), and with high ones, $M_{\rm WR}=(40$--$57)\msun$ (5 stars). The $\dot M(M_{\rm WR})$ relationship is rather steep for the former but it strongly flattens for the latter. Since {\it a posteriori\/} we find a moderate mass of the WR star in Cyg X-3, we fit only the former subsample. Using a fitting method symmetric in $M_{\rm WR}$ and $\dot M$ (see appendix C of \citealt{z11b}), we obtain
\begin{eqnarray}
\lefteqn{
\dot M\simeq \dot M_0\left(M_{\rm WR}\over \langle M\rangle\right)^n,\quad \dot M_0= (1.9\pm 0.2)\times 10^{-5} {\msun\over {\rm yr}}, \nonumber}\\
\lefteqn{
\langle M\rangle \simeq 14.7\msun,\quad n\simeq 2.93\pm 0.38,\quad M\la 22\msun,
\label{mdot_m}}
\end{eqnarray}
shown in Fig.\ \ref{mdot_vs_m}. Here, the normalization is at the average of the fitted $M_{\rm WR}$, $\langle M\rangle$. We have also fitted the subsample with $Y>0.8$ and $M_{\rm WR}<22\msun$ (with 4 less stars), but the results are very similar. On the other hand, NL00 fitted the entire WN sample, including the massive stars, and thus obtained a much flatter dependence, see their equation (23). However, it is clear from Fig.\ \ref{mdot_vs_m} that this is a very inaccurate description of the actual dependence. Equation (\ref{mdot_m}) for a fiducial $\dot M= 10^{-5}\msun$/yr implies $M_{\rm WR}\simeq (12\pm 1)\msun$. 

We note here that the fit finds the uncertainty, $\Delta\dot M_0$, of the {\it average\/} dependence. Its observed spread, or rms, is much larger, and it is partly intrinsic and partly due to measurement errors. Without detailed knowledge of the errors, we make a conservative assumption that the observed dispersion of $\dot M$ for a given $M_{\rm WR}$ is mostly intrinsic, due to the internal structure and other factors affecting $\dot M$. An estimate of the rms is $\Delta\dot M_0 \sqrt{N-1}$, where $N=29$ is the number of stars in the fitted sample. Then, in order to find the actual uncertainty range, we keep $n$ fixed and move up and down the best fit dependence within $\pm \Delta\dot M_0 \sqrt{N-1}$, as illustrated by the dotted lines in Fig.\ \ref{mdot_vs_m}. 

Note that clumping reduces $\dot M$ for a given $M_{\rm WR}$, and our relationship is a factor of 2--3 below that of Langer (1989) or \citet{lommen05}. On the other hand, our fit predicts $\dot M$ higher by a factor of 3.6 than that calculated by \citet{vdk05} for the WN 8 star WR 40 (also present in the sample used here). However, $\dot M\simeq 3.1\times 10^{-5}\msun$/yr, which is almost the same as that of NL00, was found for that star by \citet*{herald01}, who also took into account clumping. We also note that although \citet{herald01} claim that their value of $\dot M$ for another WN 8 star, WR 16, is different from that of NL00 due to a different treatment of clumping, the difference is entirely accounted for by the different values of the used distance. Summarizing, we have not found a valid objection to the results of NL00 for their WN sample.

Equations (\ref{pdot}) and (\ref{mdot_m}) can be solved for $M_{\rm WR}$ and $\dot M$ as functions of $q$, which is given by equation (\ref{ratio}),
\begin{eqnarray}
\lefteqn{
M_{\rm WR}=\left[\dot P(1+q)\langle M\rangle^n \over 2 P\dot M_0\right]^{1\over n-1},\label{mwr}}\\
\lefteqn{
\dot M  = \dot M_0^{-{1\over n-1}} \left[\dot P(1+q)\langle M\rangle \over 2 P\right]^{n\over n-1}.
\label{mdot}}
\end{eqnarray}
The $\dot M(M_{\rm WR})$ uncertainty based on the rms (see above) and the $\dot P/P$ of Kitamoto yield $M_{\rm WR}\simeq 10.3_{-2.8}^{+3.9}\msun$, $M_{\rm C}\simeq 2.4_{-1.1}^{+2.1}\msun$, $\dot M\simeq 6.5^{+3.3}_{-2.1}\times 10^{-6}\msun$/yr, and  $i={43\degr}^{+17}_{-12}$ [from equation (\ref{incl})]. The estimates based on the standard-deviation uncertainties yield somewhat narrower ranges of $M_{\rm WR}\simeq 10.3_{-1.9}^{+1.5}\msun$, $M_{\rm C}\simeq 2.4_{-1.0}^{+1.4}\msun$, $\dot M\simeq 6.5^{+1.7}_{-1.5}\times 10^{-6}\msun$/yr, and $i={43\degr}^{+14}_{-9}$. The lower limits on the masses and $\dot M$ correspond to the lower limits on $K_{\rm WR}$, $\dot P$ and $n$ and the upper limits on $K_{\rm C}$ and $\dot M_0$. Using the original mass function of V09 changes these results only slightly. Concluding, both ways of estimating the uncertainties allow either a low-mass BH or an NS to be present. We note that our mass determination is in agreement with that of \citet{tn94}, who found $M_{\rm WR}\simeq (5$--$10)\msun$ based on the observed depth of the X-ray Fe K edge and a standard stellar-wind model. 

We then compare our results on $\dot M$ with other independent estimates. The most detailed study of the constraint from X-ray absorption appears to be that of \citet{sz08}. They took into account wind clumping as well as the presence of a homogeneous wind phase. They obtained results for the cases of a low and high mass of the WR star, $6.4\msun$ and $70\msun$, respectively. Our range of $M_{\rm WR}$ is much closer to the lower one, and the dependence on $M_{\rm WR}$ is relatively modest. For the low-mass case, they obtained $\dot M\simeq (6$--$8)\times 10^{-6}\msun$/yr, almost the same as our values obtained here. A related (but independent) estimate is that of Zdziarski et al.\ (2012), who studied X-ray orbital modulation in Cyg X-3. Its orbital-phase dependence in hard X-rays, where the opacity is likely to be dominated by Thomson scattering, implies $\dot M\simeq (7$--$8)\times 10^{-6}\msun$/yr for the terminal wind velocity of 1500--1700 km/s \citep{fender99,v96}. Then, \citet{waltman96} obtained $\dot M\la 1.0\times 10^{-5}\msun$/yr from the radio delays interpreted as being due to wind opacity. This is again fully consistent with our values. The above methods are independent of the distance to Cyg X-3. 

On the other hand, \citet{v93} and \citet{ogley01} obtained higher values of $\dot M$ by interpreting the observed IR fluxes as solely due to wind emission. \citet{v93} obtained $\dot M\simeq 4\times 10^{-5}\msun$/yr without taking into account clumping of the wind. Including it lowers $\dot M$ by a factor of several \citep{sz08} and thus makes it consistent with our present results. Ogley et al.\ (2001) obtained $\dot M\simeq (4$--$30)\times 10^{-5}\msun$/yr using different assumptions about the wind, but also neglecting clumping. The lowest value is for a strongly ionized wind. Using that assumption and taking into account clumping can again reconcile their results with ours. Furthermore, the wind-emission calculations depend on the distance as $\dot M\propto d^{3/2}$, and the above authors assumed $d\simeq 10$ kpc. On the other hand, the actual distance remains relatively uncertain. \citet{p00} obtained $d\simeq 9^{+4}_{-2}$ kpc, while \citet{d83} found $d\ga 9$ kpc, but without giving an error estimate. \citet*{lzt09} gave a highly accurate result, but assuming the distance to the OB2 association of exactly 1.7 kpc. That distance has been measured as $1.74\pm 0.08$ kpc \citep{mt91}. For that range, the method of \citet{lzt09} gives $d\simeq 6$--9 kpc. If $d\simeq 7$ kpc, the values of $\dot M$ of \citet{ogley01} should be reduced by $\sim$40 per cent. In addition, there can be a substantial IR contribution from the synchrotron emission of the jets, which would further reduce the calculated $\dot M$. Summarizing, all the published estimates of $\dot M$ are compatible with our present results.

\section{A low-mass black hole?}
\label{discussion}

We have found that either an NS or a low-mass BH may be present in Cyg X-3. On the other hand, there is a lot of evidence (though still circumstantial) for the compact object in Cyg X-3 being a BH. Its spectral states have spectra showing a general resemblance to those of the canonical spectral states of BH binaries \citep*{sz08,zmg10}. The existing differences may be explained by scattering in the very strong stellar wind of Cyg X-3 \citep{zmg10}. The radio flux in the hard spectral state is strongly correlated with that in soft X-rays, which correlation is very similar \citep*{gfp03,szm08} to that seen in BH X-ray binaries. The normalization of the radio flux in Cyg X-3 is somewhat above that for BH binaries, which excess can be due to strong absorption of soft X-rays in Cyg X-3. On the other hand, NS X-ray binaries in the analogous state have the radio flux $\ga 30$ times weaker on average \citep{mf06}, in stark contrast to the case of Cyg X-3. If Cyg X-3 were indeed a progenitor of PSR B1913+16, its neutron star should be fast spinning and with moderate or high magnetic field, $\ga$ a few times $10^{10}$ G. We might expect then X-ray pulsations, which have not been found. Also, spectra expected from such a system would be rather different from those of Cyg X-3. Then, V09 (see their section 6) fitted the IR-flux orbital modulation and found it is compatible with a low inclination, $i \sim 30\degr$, but not with $i \sim 60\degr$, implying the compact object to be a BH. A low inclination, $i\simeq 30$--$40\degr$, has also been found by \citet{z12} as accounting for the orbital modulation of the X-ray continuum.

The Eddington luminosity for He accretion is $L_{\rm E}\simeq 2.5\times 10^{38}(M_{\rm C}/\msun)$ erg s$^{-1}$. The absorption corrected, phase-averaged, bolometric fluxes in the hard and soft spectral state (assuming the low-mass model of \citealt{sz08}; the hard-state flux corrected in \citealt{szm08}) are $8.5\times 10^{-9}$ erg cm$^{-2}$ s$^{-1}$ and $2.7\times 10^{-8}$ erg cm$^{-2}$ s$^{-1}$, respectively. These are orbital-phase averages; they should be increased by a factor $\simeq 1.5$ and 2 in the hard and soft state, respectively, to obtain the intrinsic values \citep{z12}. Assuming $d=7$ kpc, we obtain the bolometric hard and soft state luminosities of $\simeq 7\times 10^{37}$ erg s$^{-1}$ and $\simeq 3\times 10^{38}$ erg s$^{-1}$, respectively. At $M=4\msun$, they correspond to $\simeq 0.07L_{\rm E}$ and $\simeq 0.3L_{\rm E}$, respectively. These Eddington ratios are within the ranges observed for BH binaries in these two states \citep*{dgk07}.

Following the above circumstantial evidence and our estimate of the compact object mass ($M_{\rm C} = 2.4^{+2.1}_{-1.1} \msun$) we argue that Cyg X-3 hosts a low-mass BH. The actual BH mass is most likely in the range $(M_{\rm NS,max}$--$4.5 \msun)$, where $M_{\rm NS,max}$ is the still unknown maximum NS mass. The recent precise mass measurements of Galactic pulsars require that $M_{\rm NS,max} > 2 \msun$ \citep{demorest10}, while dynamical BH mass estimates impose that $M_{\rm NS,max} < 5 \msun$ \citep{ozel10,farr11}. Theoretical calculations of dense nuclear matter indicate that $M_{\rm NS,max} < 2.4 \msun$ \citep{lp10}. When we impose above limits on our estimate we obtain $M_{\rm BH} = 2.4^{+2.1}_{-0.4} \msun$. This puts the BH in Cyg X-3 right into the mass gap, the dearth of compact objects in the (2--$5) \msun$ mass range, first noted by \citet{bailyn98}.

The existence of the mass gap was recently discussed in context of supernova explosion models \citep{belczynski12}. For the rapidly developing explosions (within $\sim$0.2 s after the core bounce), the mass gap appears in simulations as observed for the Galactic BH X-ray binaries. However, there is a small addition of systems with compact objects in the (2--$3.5) \msun$ mass range (see fig.~1 of \citealt{belczynski12}). These have formed via accretion onto an NS (originally formed in the supernova explosion) from a binary companion. These compact objects are most likely light BHs formed via accretion-induced collapse of the NS \citep{vdl73,bb98}. We note, however, that the result of \citet{bb98} that an NS can accrete as much as $\sim 1\msun$ was questioned by \citet*{bkb02}, who found that usually a much smaller mass can be accreted. The numerical calculations of \citet{ruffert99} and \citet{rt08} confirm that the accretion rate onto compact objects in common-envelope events can be significantly lower (by a factor of $\sim 10$) than the Bondi-Hoyle rate.

The mass gap was also recently interpreted as a potential observational artefact caused by the systematic uncertainties in dynamical BH mass measurements \citep{kreidberg12}. No mass gap scenario corresponds to delayed supernova models (developing on a timescale $\sim$0.5--1 s) that produce compact objects within (2--$5) \msun$ mass range \citep{belczynski12}. In this case, the light BH in Cyg X-3 most likely have formed in supernova explosion from an $M_{\rm zams} \simeq (20$--$40) \msun$ progenitor \citep{fryer12}. As long as there is no consensus on the mass gap existence, both formation scenarios, either via accretion induced collapse or in the delayed supernova explosion, are to be considered.

Finally, we mention that we can expect a large number of He binaries in the Galaxy with the donor masses lower than that in Cyg X-3 (and containing either NS or BH). Given our obtained low to moderate WR mass in Cyg X-3, many of them would have masses $\la 6\msun$, where the He stars would be no more of the WR type due to the weakness of their wind. These systems would then be undetectable as X-ray sources. As pointed out by \citet{vdh95}, this would help to explain the existence of only one system like Cyg X-3 in our Galaxy.

Summarizing, we have calculated the masses of the components, the inclination and the donor mass-loss rate in the Cyg X-3 binary. We have used existing mass functions, the relationship between $\dot M$ and $\dot P$ and our new determination of the dependence of the mass-loss rate on the mass for WN stars of moderate mass. The obtained $\dot M$ is in a very good agreement with two independent calculations, and are consistent with all other determinations, provided those based on the IR fluxes are corrected for clumping. We argue that although an NS is allowed to be present, a low-mass BH is more likely given the radio, IR and X-ray properties of Cyg X-3. In the companion paper \citep{paper2}, we study the fate of this system, which is likely to first form either an BH-NS or BH-BH binary, and then merge, giving rise to a detectable gravitational-wave signal. 

\section*{ACKNOWLEDGMENTS}
This research has been supported in part by the Polish NCN grants N N203 581240, N N203 404939, 2012/04/M/ST9/00780, and the NASA Grant NNX09AV06A to the UTB Center for Gravitational Wave Astronomy (KB). We thank Janusz Zi{\'o}{\l}kowski for a valuable discussion, and the referee for valuable suggestions.

\label{lastpage}

\begin{thebibliography}{}

\bibitem[\protect\citeauthoryear{Abdo et al.}{2009}]{fermi09} 
Abdo A. A., et al., 2009, Sci, 326, 1512

\bibitem[\protect\citeauthoryear{Bailyn et al.}{1998}]{bailyn98} 
Bailyn C.~D., Jain R.~K., Coppi P., Orosz J.~A., 1998, ApJ, 499, 367 

\bibitem[\protect\citeauthoryear{Belczy{\'n}ski, Kalogera \& Bulik}{Belczy{\'n}ski et al.}{2002}]{bkb02} 
Belczy{\'n}ski K., Kalogera V., Bulik T., 2002, ApJ, 572, 407 

\bibitem[\protect\citeauthoryear{Belczy{\'n}ski et al.}{2012a}]{paper2} 
Belczy{\'n}ski K., Bulik T., Mandel I., Sathyaprakash B.~S., Zdziarski A. A., Miko{\l}ajewska J., 2012a, ApJ, submitted, arXiv:1209.2658 

\bibitem[\protect\citeauthoryear{Belczy{\'n}ski et al.}{2012b}]{belczynski12} 
Belczy{\'n}ski K., Wiktorowicz G., Fryer C., Holz D., Kalogera V., 2012b, ApJ, 757, 91

\bibitem[\protect\citeauthoryear{Bethe \& Brown}{1998}]{bb98} 
Bethe H.~A., Brown G.~E., 1998, ApJ, 506, 780 

\bibitem[\protect\citeauthoryear{Davidsen \& Ostriker}{1974}]{do74} 
Davidsen A., Ostriker J.~P., 1974, ApJ, 189, 331 

\bibitem[\protect\citeauthoryear{Demorest et al.}{2010}]{demorest10} 
Demorest P.~B., Pennucci T., Ransom S.~M., Roberts M.~S.~E., Hessels J.~W.~T., 2010, Natur, 467, 1081 

\bibitem[\protect\citeauthoryear{Dickey}{1983}]{d83} 
Dickey J. M., 1983, ApJ, 273, L71

\bibitem[\protect\citeauthoryear{Done, Gierli{\'n}ski \& Kubota}{Done et al.}{2007}]{dgk07} 
Done C., Gierli{\'n}ski M., Kubota A., 2007, A\&ARv, 15, 1 

\bibitem[\protect\citeauthoryear{Farr et al.}{2011}]{farr11} 
Farr W.~M., Sravan N., Cantrell A., Kreidberg L., Bailyn C.~D., Mandel I., Kalogera V., 2011, ApJ, 741, 103 

\bibitem[\protect\citeauthoryear{Fender, Hanson \& Pooley}{Fender et al.}{1999}]{fender99} 
Fender R.~P., Hanson M.~M., Pooley G.~G., 1999, MNRAS, 308, 473 

\bibitem[\protect\citeauthoryear{Fryer et al.}{2012}]{fryer12} 
Fryer C.~L., Belczy{\'n}ski K., Wiktorowicz G., Dominik M., Kalogera V., Holz 
D.~E., 2012, ApJ, 749, 91 

\bibitem[\protect\citeauthoryear{Gallo, Fender \& Pooley}{Gallo et al.}{2003}]{gfp03}
Gallo E., Fender R.~P., Pooley G.~G., 2003, MNRAS, 344, 60

\bibitem[\protect\citeauthoryear{Giacconi et al.}{1967}]{giacconi67} 
Giacconi R., Gorenstein P., Gursky H., Waters J. R., 1967, ApJ, 148, L119

\bibitem[\protect\citeauthoryear{Hanson, Still \& Fender}{Hanson et al.}{2000}]{hanson00} 
Hanson M.~M., Still M.~D., Fender R.~P., 2000, ApJ, 541, 308 

\bibitem[\protect\citeauthoryear{Herald, Hillier \& Schulte-Ladbeck}{Herald et al.}{2001}]{herald01} 
Herald J.~E., Hillier D.~J., Schulte-Ladbeck R.~E., 2001, ApJ, 548, 932 

\bibitem[\protect\citeauthoryear{Hulse \& Taylor}{1975}]{ht75} 
Hulse R.~A., Taylor J.~H., 1975, ApJ, 195, L51 

\bibitem[\protect\citeauthoryear{Kreidberg et al.}{2012}]{kreidberg12} 
Kreidberg L., Bailyn C., Farr W., Kalogera V. 2012, ApJ, 757, 36

\bibitem[\protect\citeauthoryear{Langer}{1989}]{langer89} 
Langer N., 1989, A\&A, 220, 135 

\bibitem[\protect\citeauthoryear{Lattimer \& Prakash}{2010}]{lp10} 
Lattimer J.~M., Prakash M., 2010, In Lee S., ed., Gerry Brown's Festschrift. World Scientific, arXiv:1012.3208 

\bibitem[\protect\citeauthoryear{Ling, Zhang \& Tang}{Ling et al.}{2009}]{lzt09} 
Ling Z., Zhang S. \& Tang S., 2009, ApJ, 695, 1111

\bibitem[\protect\citeauthoryear{Lommen et al.}{2005}]{lommen05} 
Lommen D., Yungelson L., van den Heuvel E., Nelemans G., Portegies Zwart S., 2005, A\&A, 443, 231

\bibitem[\protect\citeauthoryear{Mart{\'{\i}}, Paredes \& Peracaula}{Marti et al.}{2001}]{marti01} 
Mart{\'{\i}} J., Paredes J.~M., Peracaula M., 2001, A\&A, 375, 476 

\bibitem[\protect\citeauthoryear{Massey \& Thompson}{1991}]{mt91} 
Massey P., Thompson A.~B., 1991, AJ, 101, 1408 

\bibitem[\protect\citeauthoryear{McCollough et al.}{1999}]{mccollough99} 
McCollough M. L., et al.\ 1999, ApJ, 517, 951

\bibitem[\protect\citeauthoryear{Migliari \& Fender}{2006}]{mf06} 
Migliari S., Fender R.~P., 2006, MNRAS, 366, 79 

\bibitem[\protect\citeauthoryear{Mioduszewski et al.}{2001}]{m01} 
Mioduszewski A.~J., Rupen M.~P., Hjellming R.~M., Pooley G.~G., Waltman E.~B., 2001, ApJ, 553, 766 

\bibitem[\protect\citeauthoryear{Nugis \& Lamers}{2000}]{nl00}
Nugis T., Lamers H. J. G. L. M., 2000, A\&A, 360, 227 (NL00)

\bibitem[\protect\citeauthoryear{Ogley, Bell Burnell \& Fender}{Ogley et al.}{2001}]{ogley01} 
Ogley R.~N., Bell Burnell S.~J., Fender R.~P., 2001, MNRAS, 322, 177 

\bibitem[\protect\citeauthoryear{{\"O}zel et al.}{2010}]{ozel10} 
{\"O}zel F., Psaltis D., Narayan R., McClintock J.~E., 2010, ApJ, 725, 1918 

\bibitem[\protect\citeauthoryear{Paczy\'nski}{1971}]{p71}
Paczy\'nski B., 1971, ARA\&A, 9, 183

\bibitem[\protect\citeauthoryear{Predehl et al.}{2000}]{p00}
Predehl P., Burwitz V., Paerels F., Tr{\"u}mper J., 2000, A\&A, 357, L25 

\bibitem[\protect\citeauthoryear{Ricker \& Taam}{2008}]{rt08} 
Ricker P. M., Taam R. E., 2008, ApJ, 672, L41

\bibitem[\protect\citeauthoryear{Ruffert}{1999}]{ruffert99} 
Ruffert M., 1999, A\&A, 346, 861

\bibitem[\protect\citeauthoryear{Schaerer \& Maeder}{1992}]{sm92} 
Schaerer D., Maeder A., 1992, A\&A, 263, 129 

\bibitem[\protect\citeauthoryear{Singh et al.}{2002}]{singh02} 
Singh N.~S., Naik S., Paul B., Agrawal P.~C., Rao A.~R., Singh K.~Y., 2002, A\&A, 392, 161 

\bibitem[\protect\citeauthoryear{Szostek \& Zdziarski}{2008}]{sz08} 
Szostek A., Zdziarski A. A., 2008, MNRAS, 386, 593

\bibitem[\protect\citeauthoryear{Szostek, Zdziarski \& McCollough}{Szostek et al.}{2008}]{szm08} 
Szostek A., Zdziarski A.~A., McCollough M.~L., 2008, MNRAS, 388, 1001

\bibitem[\protect\citeauthoryear{Tavani et al.}{2009}]{agile} 
Tavani M., et al., 2009, Nat, 462, 620 

\bibitem[\protect\citeauthoryear{Terasawa \& Nakamura}{1994}]{tn94} 
Terasawa N., Nakamura H., 1994, ApJS, 92, 477 

\bibitem[\protect\citeauthoryear{van den Heuvel}{1995}]{vdh95} 
van den Heuvel E.~P.~J., 1995, J.\ Astrophys.\ Astr., 16, 255 

\bibitem[\protect\citeauthoryear{van den Heuvel \& De Loore}{1973}]{vdl73} 
van den Heuvel E.~P.~J., De Loore C., 1973, A\&A, 25, 387 

\bibitem[\protect\citeauthoryear{van Kerkwijk}{1993}]{v93} 
van Kerkwijk M.~H., 1993, A\&A, 276, L9 

\bibitem[\protect\citeauthoryear{van Kerkwijk et al.}{1992}]{v92} 
van Kerkwijk M.~H., et al., 1992, Nat, 355, 703 

\bibitem[\protect\citeauthoryear{van Kerkwijk et al.}{1996}]{v96} 
van Kerkwijk M. H., Geballe T. R., King D. L., van der Klis M.,
van Paradijs J., 1996, A\&A, 314, 521

\bibitem[\protect\citeauthoryear{Vilhu et al.}{2009}]{vilhu09} 
Vilhu O., Hakala P., Hannikainen D.~C., McCollough M., Koljonen K., 2009, A\&A, 501, 679 (V09)

\bibitem[\protect\citeauthoryear{Vink \& de Koter}{2005}]{vdk05} 
Vink J., de Koter A., 2005, A\&A, 442, 587

\bibitem[\protect\citeauthoryear{Waltman et al.}{1996}]{waltman96}
Waltman E. B., Foster R. S., Pooley G. G., Fender R. P., Ghigo F. D., 1996, AJ, 112, 2690

\bibitem[\protect\citeauthoryear{Zdziarski, Misra \& Gierli{\'n}ski}{Zdziarski et al.}{2010}]{zmg10} 
Zdziarski A.~A., Misra R., Gierli{\'n}ski M., 2010, MNRAS, 402, 767 

\bibitem[\protect\citeauthoryear{Zdziarski et al.}{2011}]{z11b} 
Zdziarski A.~A., Skinner G.~K., Pooley G.~G., Lubi{\'n}ski P., 2011, MNRAS, 416, 1324

\bibitem[\protect\citeauthoryear{Zdziarski et al.}{2012}]{z12} 
Zdziarski A.~A., Maitra C., Frankowski A., Skinner G.~K., Misra R., 2012, MNRAS, 426, 1031

\end{thebibliography}
\end{document}